\documentclass[usegraphicx]{mn2e}

\title[A new set of tight scaling relations]
      {From massive spirals to dwarf irregulars:
       a new set of tight scaling relations for cold gas and stars driven by
       disc gravitational instability}

\author[A. B. Romeo]
       {Alessandro B. Romeo$^{1}$\thanks{E-mail: romeo@chalmers.se}\\
        $^{1}$Department of Space, Earth and Environment,
              Chalmers University of Technology,
              SE-41296 Gothenburg, Sweden}

\begin{document}

\date{Accepted 2019 November 28.
      Received 2019 November 25; in original form 2019 May 14}

\pagerange{\pageref{firstpage}--\pageref{lastpage}}

\pubyear{2020}

\maketitle

\label{firstpage}

\begin{abstract}
We present a new set of galaxy scaling relations for the relative mass
content of atomic gas, molecular gas and stars.  Such relations are driven by
disc gravitational instability, and originate from the low galaxy-to-galaxy
variance of Toomre's $Q$ stability parameter.  We test such relations using
more than 100 galaxies, from massive spirals to dwarf irregulars, thus
spanning several orders of magnitude in stellar mass
($M_{\star}\approx10^{6\mbox{--}11}\,\mbox{M}_{\odot}$) and atomic gas mass
($M_{\mathrm{HI}}\approx10^{7\mbox{--}10.5}\,\mbox{M}_{\odot}$).  Such tests
demonstrate (i) that our scaling relations are physically motivated and
tightly constrained, (ii) that the mass-averaged gravitational instability
properties of galaxy discs are remarkably uniform across the sequence
Sa--dIrr, and (iii) that specific angular momentum plays an important role in
such a scenario.  Besides providing new insights into a very important topic
in galaxy evolution, this work provides a simple formula (Eq.\ 5) that one
can use for generating other galaxy relations driven by disc instability.  We
explain how to do that, mention a few possible applications, and stress the
importance of testing our approach further.
\end{abstract}


\section{INTRODUCTION}

Statistical correlations between physical properties of galaxies are
indispensable tools for unravelling the fundamental laws that govern galaxy
formation and evolution across the observed variety of scales.  Such `scaling
relations' are therefore constantly used for testing simulation and
semi-analytic models of galaxy evolution, and for constraining their
predictions (e.g., Dutton et al.\ 2011; Lagos et al.\ 2016; Agertz et
al.\ 2019; Forbes et al.\ 2019; Ginolfi et al.\ 2019).  Discovering new
scaling relations is thus very important, and so is determining the physical
processes that drive them.

Recent examples of galaxy scaling relations that have attracted special
interest are those linking the relative mass content of atomic and molecular
gas to stellar mass, or to related properties like stellar mass surface
density, specific star formation rate and colour (e.g., Saintonge et
al.\ 2011; Huang et al.\ 2012; Boselli et al.\ 2014; Catinella et al.\ 2018).
Recent investigations focusing on spiral galaxies suggest that the atomic gas
scaling relation is driven by disc gravitational instability.  Obreschkow et
al.\ (2016) proposed a hybrid stability model that predicts the mass fraction
of atomic (hydrogen+helium) gas as a function of mass and specific angular
momentum of the whole (gas+stars) disc via a newly defined $q$ parameter,
assuming a constant H\,\textsc{i} velocity dispersion
($\sigma_{\mathrm{HI}}=10\;\mbox{km\,s}^{-1}$).  Such a stability model has
been tested in a variety of applications, and found to be a reliable atomic
gas tracer (e.g., Lagos et al.\ 2017; Lutz et al.\ 2017, 2018; Stevens et
al.\ 2018; Wang et al.\ 2018; D\v{z}ud\v{z}ar et al.\ 2019; Murugeshan et
al.\ 2019; Stevens et al.\ 2019).  Another important contribution is the one
by Zasov \& Zaitseva (2017), who showed that the relation between atomic gas
mass and disc specific angular momentum is equally well described by a
simpler stability model controlled by $Q_{\mathrm{gas}}$, the gas Toomre
parameter, assuming that $Q_{\mathrm{gas}}$ is approximately constant within
a galaxy (like $\sigma_{\mathrm{gas}}$).  This stability model was tested and
constrained by Kurapati et al.\ (2018).

The above-mentioned link between disc gravitational instability and disc
angular momentum is another important aspect of the problem and has been
investigated in many other works, not only in the original context of
instability to bar formation (e.g., Mo et al.\ 1998; Athanassoula 2008;
Agertz \& Kravtsov 2016; Sellwood 2016; Okamura et al.\ 2018; Romeo \&
Mogotsi 2018; Zoldan et al.\ 2018; Valencia-Enr\'{i}quez et al.\ 2019) but
also in the context of Toomre instability, both at low $z$ (e.g., Obreschkow
\& Glazebrook 2014; Stevens et al.\ 2016; Romeo \& Mogotsi 2018) and at high
$z$ (e.g., Obreschkow et al.\ 2015; Stevens et al.\ 2016; Swinbank et
al.\ 2017; Sweet et al.\ 2019).  All such works find that there is a
measurable link between disc gravitational instability and galactic angular
momentum, but the resulting relation depends on which stability diagnostic is
used.

Romeo \& Mogotsi (2018) pointed out that $Q_{\mathrm{gas}}$ and $q$
(Obreschkow et al.\ 2016) are not fully reliable stability diagnostics.  This
concerns not only nearby spirals, where stars are the primary driver of disc
instabilities (Romeo \& Fathi 2016; Romeo \& Mogotsi 2017), but also gas-rich
galaxies at high $z$ and their nearby analogues, where turbulence can drive
the disc into regimes that are far from Toomre/Jeans instability (Romeo et
al.\ 2010; Romeo \& Agertz 2014; see also Renaud 2019).  Romeo \& Mogotsi
(2018) showed that a more reliable stability diagnostic for spiral galaxies
is $\langle\mathcal{Q}_{\star}\rangle$, the mass-weighted average of the
stellar Toomre parameter, corrected so as to include disc thickness effects
(Romeo \& Falstad 2013).  Such a diagnostic allowed them to tightly constrain
the relation between stellar mass, stellar specific angular momentum and disc
stability level.  Romeo \& Mogotsi (2018) also showed that
$\langle\mathcal{Q}_{\star}\rangle$ is related to the
Efstathiou-Lake-Negroponte (1982) global stability parameter,
$\epsilon_{\mathrm{m}}$, via the degree of rotational support, $V/\sigma$,
and the velocity dispersion anisotropy, $\sigma_{z}/\sigma_{R}$.  These are
two important effects that are missing from $\epsilon_{\mathrm{m}}$.

This paper proposes a unified approach to the problem, which results in a new
set of scaling relations for the relative mass content of cold gas and stars,
and which is able to generate further useful scaling relations, all driven by
disc gravitational instability.  Unlike the models of Obreschkow et
al.\ (2016) and Zasov \& Zaitseva (2017), our approach does not assume that
$\sigma_{\mathrm{gas}}$ or $Q_{\mathrm{gas}}$ is approximately constant
within a galaxy, and applies not only to atomic gas but also to molecular gas
and stars.  While the present approach builds on the analyses performed by
Romeo \& Mogotsi (2017, 2018), the applications presented here are entirely
new and concern disc-dominated galaxies of all morphological types
(Sa--dIrr), thus spanning several orders of magnitude in stellar mass
($M_{\star}\approx10^{6\mbox{--}11}\,\mbox{M}_{\odot}$) and atomic gas mass
($M_{\mathrm{HI}}\approx10^{7\mbox{--}10.5}\,\mbox{M}_{\odot}$).  Our
approach is presented in Sect.\ 2, where we show the route that connects
Toomre's $Q$ stability parameter to our new set of scaling relations, and
beyond.  Galaxy samples, data and statistics are described in Sect.\ 3, the
new set of scaling relations is presented and tested in Sect.\ 4, and the
conclusions are drawn in Sect.\ 5.

\section{A UNIFIED APPROACH FOR COLD GAS AND STARS}

\subsection{Choice of a representative galaxy sample and basic quantities}

As a representative galaxy sample, we choose the popular sample of spirals
originally selected and analysed by Leroy et al.\ (2008), hereafter L08: NGC
628, 2841, 3184, 3198, 3351, 3521, 3627, 4736, 5055, 5194, 6946 and 7331.
These are 12 nearby star-forming galaxies with sensitive and spatially
resolved measurements of atomic gas, molecular gas and stellar properties
across the entire optical disc, compiled by combining data from several
surveys.  We refer to L08 for a detailed description of the data and their
translation into physical quantities (see their sect.\ 3).

L08's sample of spirals provides all the quantities needed to carry out our
analysis.  In particular, among other quantities specified in Sect.\ 2.2, we
use the same epicyclic frequency ($\kappa$), mass surface densities of atomic
gas ($\Sigma_{\mathrm{HI}}$), molecular gas ($\Sigma_{\mathrm{H2}}$) and
stars ($\Sigma_{\star}$), and stellar radial velocity dispersion
($\sigma_{\star}$) as in L08 (see their appendices A--C and E--F).  However,
rather than using observationally motivated values of the H\,\textsc{i} and
$\mbox{H}_{2}$ velocity dispersions, we use observed radial profiles of
$\sigma_{\mathrm{HI}}$ and $\sigma_{\mathrm{H2}}$, which rise towards the
centre in most of the galaxies (Romeo \& Mogotsi 2017; see their sect.\ 2.1,
and figs 1 and 2).  Note also that the radial profiles of $\sigma_{\star}$
derived by L08 are not based on observations, but on a simple model that
relates $\sigma_{\star}$ to the mass surface density and scale length of the
stellar disc (see their appendix B.3).  To the best of our knowledge, stellar
velocity dispersions have only been measured in three galaxies of the sample:
NGC 628 (Ganda et al.\ 2006; Herrmann \& Ciardullo 2009), NGC 3198 (Bottema
1988, 1993) and NGC 4736 (Herrmann \& Ciardullo 2009).  However, L08's model
is quite reliable.  Systematic uncertainties in $\sigma_{\star}$ are
significant, on average, only in the innermost/outermost regions of the
stellar disc, where this model overestimates/underestimates the observed
$\sigma_{\star}$ (Romeo \& Mogotsi 2017; Mogotsi \& Romeo 2019).  Those
regions, however, have little weight in the present analysis.

\subsection{Tour through disc gravitational instability}

%%%%%%%%%%%%%%%%%%%%%%%%%%%%%%%%%%%%%%%%%%%%%%%%%%%%%%%%%%%%%%%%%%%%%%%%%%%%%
\begin{figure*}
\includegraphics[scale=1.16]{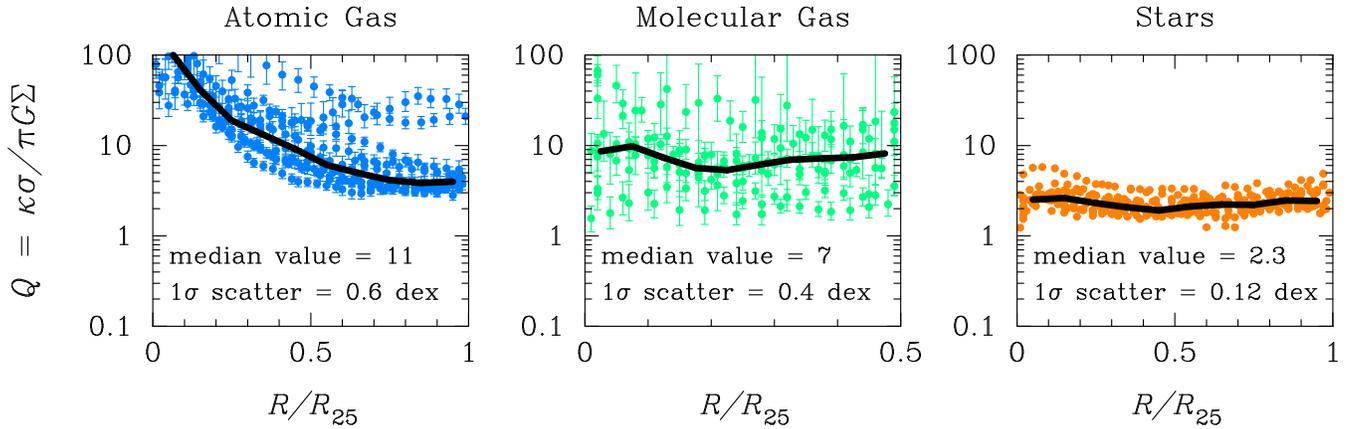}
\caption{Radial profiles of the Toomre parameter for L08's sample of spirals,
  with the galactocentric distance measured in units of the optical radius
  (B-band isophotal radius at 25 mag arcsec$^{-2}$).  Also shown is the local
  median of $Q$.  In the case of molecular gas, the radial range is limited
  by the sparsity of sensitive CO measurements beyond half the optical
  radius.}
\end{figure*}
%%%%%%%%%%%%%%%%%%%%%%%%%%%%%%%%%%%%%%%%%%%%%%%%%%%%%%%%%%%%%%%%%%%%%%%%%%%%%

To explore the link between disc gravitational instability and the relative
mass content of atomic gas, molecular gas and stars in galaxies, we start
from the simplest stability diagnostic: the Toomre (1964) parameter,
$Q=\kappa\sigma/\pi G\Sigma$.  It is commonly assumed that $Q\approx1$,
consistent with a process of self-regulation that keeps galaxy discs close to
marginal stability (see sect.\ 1 of Krumholz et al.\ 2018 for an overview).
How realistic is that assumption?  Fig.\ 1 illustrates that atomic gas,
molecular gas and stars have distinct radial distributions of $Q$, which
differ both in median trend and in variance.  While $Q_{\star}$ is quite
close to unity, $Q_{\mathrm{H2}}$ is three times more offset and scattered,
whereas $Q_{\mathrm{HI}}$ exhibits a two-orders-of-magnitude decline within
the optical radius and an even larger median offset from unity than
$Q_{\mathrm{H2}}$.  Thus the assumption that $Q\approx1$ is not realistic
enough to represent the diverse phenomenology of $Q$ in galaxy discs.

Such a diversity results from the complex interplay between the heating and
cooling processes that regulate the value of $Q$ in galaxy discs (Krumholz \&
Burkert 2010; Forbes et al.\ 2012, 2014, 2019), and from the fact that
$Q_{\mathrm{HI}}$, $Q_{\mathrm{H2}}$ and $Q_{\star}$ do not really measure
the stability levels of atomic gas, molecular gas and stars.
$Q_{\mathrm{HI}}$, $Q_{\mathrm{H2}}$ and $Q_{\star}$ are instead the building
blocks of a more realistic, multi-component $\mathcal{Q}$ stability parameter
(Romeo \& Falstad 2013).  Such a parameter is dominated by $Q_{\star}$
because stars, and not molecular or atomic gas, are the primary driver of
disc instabilities in spiral galaxies (Romeo \& Mogotsi 2017; Marchuk 2018;
Marchuk \& Sotnikova 2018; Mogotsi \& Romeo 2019), which is true even for a
powerful starburst+Seyfert galaxy like NGC 1068 (Romeo \& Fathi 2016).  This
is the reason why $Q_{\star}$ is much closer to the critical stability level
($\mathcal{Q}_{\mathrm{crit}}\approx2\mbox{--}3$) than $Q_{\mathrm{H2}}$ or
$Q_{\mathrm{HI}}$.  Note that $\mathcal{Q}_{\mathrm{crit}}$ is higher than
unity, but its precise value is still questioned (Romeo \& Fathi 2015).  In
fact, $\mathcal{Q}_{\mathrm{crit}}$ is influenced by complex phenomena such
as non-axisymmetric perturbations (e.g., Griv \& Gedalin 2012) and gas
dissipation (Elmegreen 2011), whose effects are difficult to quantify.  Disc
thickness effects are instead easier to evaluate, and are already included in
the definition of $\mathcal{Q}$ (see again Romeo \& Falstad 2013).

Now that we have clarified how self-regulated galaxy discs are, let us
analyse how the Toomre parameter of component $i$,
$Q_{i}=\kappa\sigma_{i}/\pi G\Sigma_{i}$, varies from galaxy to galaxy.  To
suppress the variation of $Q_{i}$ within a galaxy, we take the mass-weighted
average of $Q_{i}(R)$:
\begin{equation}
\langle Q_{i}\rangle=
\frac{1}{M_{i}(R_{\mathrm{av}})}
\int_{0}^{R_{\mathrm{av}}}Q_{i}(R)\,\Sigma_{i}(R)\,2\pi R\,\mathrm{d}R\,.
\end{equation}
This type of average is especially useful because it relates $\langle
Q_{i}\rangle$ to fundamental galaxy properties such as mass, $M_{i}$, and
specific angular momentum, $j_{i}=J_{i}/M_{i}$, via a simple and accurate
approximation (Romeo \& Mogotsi 2018).  The resulting relation is $\langle
Q_{i}\rangle\propto\mathcal{A}_{i}$, where
\begin{equation}
\mathcal{A}_{i}=
\frac{j_{i}\overline{\sigma}_{i}}{GM_{i}}\,,
\end{equation}
\begin{equation}
j_{i}=
\frac{1}{M_{i}}
\int_{0}^{\infty}Rv_{\mathrm{c}}(R)\,\Sigma_{i}(R)\,2\pi R\,\mathrm{d}R\,,
\end{equation}
\begin{equation}
\overline{\sigma}_{i}=
\frac{1}{R_{\mathrm{av}}}
\int_{0}^{R_{\mathrm{av}}}\sigma_{i}(R)\,\mathrm{d}R\,.
\end{equation}
Note four points concerning Eqs (1)--(4):
\begin{itemize}
\item $M_{i}$ and $j_{i}$ are the total mass and the total specific angular
  momentum of atomic hydrogen+helium gas ($i=\mbox{H\,\textsc{i}}$),
  molecular hydrogen+helium gas ($i=\mbox{H}_{2}$) or stars ($i=\star$).
\item Our definition of $j_{i}$ is based on that of Obreschkow \& Glazebrook
  (2014), and assumes that stars and gas follow exactly the same rotation
  curve.  This is technically not correct and tends to overestimate
  $j_{\star}$ (El-Badry et al.\ 2018; Fall \& Romanowsky 2018; Posti et
  al.\ 2018a) because it neglects asymmetric drift corrections, which are
  significant where $\sigma_{\star}\ga v_{\mathrm{c}}$ (see, e.g., Binney \&
  Tremaine 2008).  However, our definition of $j_{i}$ is the one most
  commonly used for precision measurement of angular momentum in disc
  galaxies (e.g., Obreschkow \& Glazebrook 2014; Butler et al.\ 2017;
  Chowdhury \& Chengalur 2017; Elson 2017; Kurapati et al.\ 2018).  More
  importantly, our definition of $j_{i}$ is fully consistent with the
  epicyclic approximation, $\sigma_{i}/R\kappa\ll1$ (hence $\sigma_{i}\ll
  v_{\mathrm{c}}$), a fundamental assumption behind Toomre's stability
  criterion and its descendants (see, e.g., Bertin 2014).
\item While $\langle Q_{i}\rangle$ is the mass-weighted average of
  $Q_{i}(R)$, $\overline{\sigma}_{i}$ is the radial average of
  $\sigma_{i}(R)$, where $\sigma$ denotes the radial velocity dispersion.
\item $\mathcal{A}_{i}$ may look identical to the $q$ parameter defined by
  Obreschkow et al.\ (2016), but it is not.  For instance, $\mathcal{A}_{i}$
  applies not only to atomic gas but also to molecular gas and stars, and
  does not assume that $\sigma_{i}$ is constant.  As pointed out in Sect.\ 1,
  our approach is also significantly different from the Obreschkow et
  al.\ (2016) model, and so are the resulting scaling relations, as we will
  show in Sect.\ 4.2.
\end{itemize}
Last but not least, note that the coefficient of proportionality between
$\langle Q_{i}\rangle$ and $\mathcal{A}_{i}$ is a numerical factor that
depends on $R_{\mathrm{av}}/l_{i}$, where $R_{\mathrm{av}}$ is the radius
over which $Q_{i}(R)$ is averaged (see Eq.\ 1) and $l_{i}$ is the exponential
scale length of component $i$.  So this factor is not well defined for a
component whose mass distribution is far from exponential, like atomic gas
(e.g., Bigiel \& Blitz 2012).  In view of that, we opt for a unified approach
and use $\mathcal{A}_{i}$ as a proxy for $\langle Q_{i}\rangle$: it is well
defined for all the components, and it is simpler than $\langle
Q_{i}\rangle$.  In addition, the offset of $\mathcal{A}_{i}$ from $\langle
Q_{i}\rangle$ is not an issue because $Q_{i}=1$ no longer means marginal
stability when the disc has multiple, gravitationally coupled components
(Romeo \& Falstad 2013; see also Fig.\ 1 and its discussion), and because any
such numerical factor will be statistically suppressed by the final rescaling
made in Sect.\ 2.3.  This is also the reason why we have not corrected
$Q_{i}$ so as to include disc thickness effects.

%%%%%%%%%%%%%%%%%%%%%%%%%%%%%%%%%%%%%%%%%%%%%%%%%%%%%%%%%%%%%%%%%%%%%%%%%%%%%
\begin{figure*}
\includegraphics[scale=1.16]{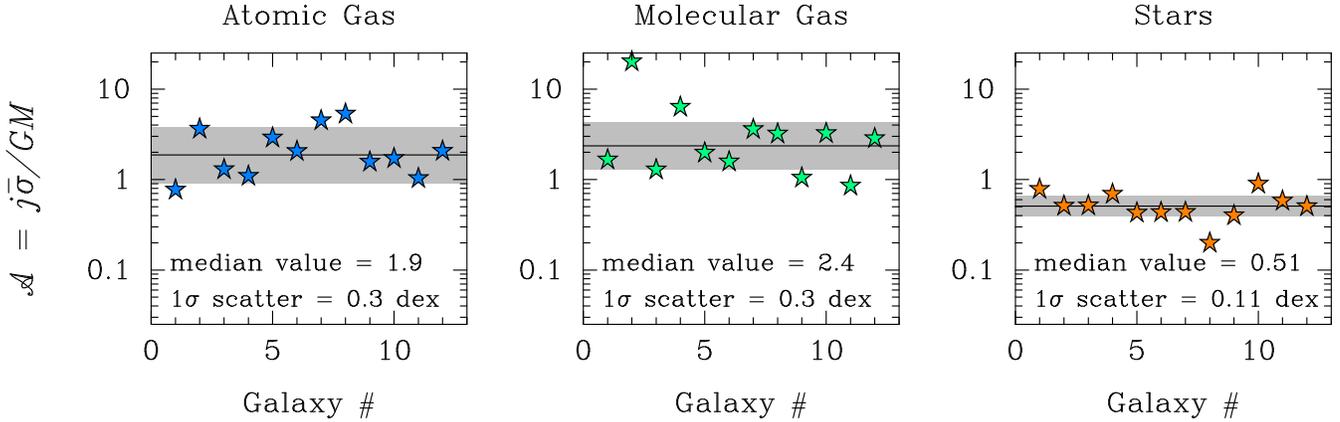}
\caption{Galaxy-to-galaxy variation of the $\mathcal{A}$ stability parameter
  for L08's sample of spirals (the galaxy list is given in the first
  paragraph of Sect.\ 2.1).  Also shown are the median value and $1\sigma$
  scatter of $\mathcal{A}$.}
\end{figure*}
%%%%%%%%%%%%%%%%%%%%%%%%%%%%%%%%%%%%%%%%%%%%%%%%%%%%%%%%%%%%%%%%%%%%%%%%%%%%%

To compute the $\mathcal{A}_{i}$ stability parameter for L08's sample of
spirals, we use the values of $M_{i}$ and $j_{i}$ tabulated by L08 and
Obreschkow \& Glazebrook (2014), respectively.  We also need to evaluate
$\overline{\sigma}_{i}$, hence to choose the averaging radius
$R_{\mathrm{av}}$.  Although one can do that arbitrarily, we prefer to make
use of all the information provided by the $\sigma_{i}$ measurements.
Therefore we choose $R_{\mathrm{av}}=R_{25}$ for atomic gas,
$R_{\mathrm{av}}=\frac{1}{2}R_{25}$ for molecular gas and
$R_{\mathrm{av}}=R_{25}$ for stars, where $R_{25}$ is the optical radius
(B-band isophotal radius at 25 mag arcsec$^{-2}$).  In the case of molecular
gas, the radial range is limited by the sparsity of sensitive CO measurements
beyond half the optical radius (see fig.\ 1 of Romeo \& Mogotsi 2017 and its
discussion).  Fig.\ 2 shows that the $1\sigma$ scatter of $\mathcal{A}_{i}$
ranges from 0.1 dex, the value measured for stars, to 0.3 dex, the value
measured for molecular and atomic gas.  We have also carried out various
tests, which show that the $1\sigma$ scatter of $\mathcal{A}_{i}$ is
unaffected by the choice of $R_{\mathrm{av}}$, even if $R_{\mathrm{av}}$ is
as small as $0.3\,R_{25}$.  Hereafter we will denote such galaxy-to-galaxy
scatter with $\sigma_{\mathrm{gg}}(i)$.%
\footnote{$\sigma_{\mathrm{gg}}(i)$ can be combined with the total scatter
  given in Fig.\ 1, $\sigma_{\mathrm{tot}}(i)$, to estimate the rms scatter
  of $Q_{i}$ within a galaxy: $\sigma_{\mathrm{g}}(i)=
  \sqrt{\sigma_{\mathrm{tot}}^{2}(i)-\sigma_{\mathrm{gg}}^2(i)}\,$.  Hence
  $\sigma_{\mathrm{g}}(\mbox{H\,\textsc{i}})=\mbox{0.5 dex}$,
  $\sigma_{\mathrm{g}}(\mbox{H}_{2})=\mbox{0.2 dex}$ and
  $\sigma_{\mathrm{g}}(\star)=\mbox{0.05 dex}$.}

Before focusing on a more important meaning of $\sigma_{\mathrm{gg}}$ (see
Sect.\ 2.3), let us test the robustness of the result
$\sigma_{\mathrm{gg}}(\star)<\sigma_{\mathrm{gg}}(\mbox{gas})$ further.  Is
this an artifact of L08's model-based $\sigma_{\star}(R)$?  Current
integral-field-unit (IFU) surveys allow measuring the mass-weighted average
of $Q_{\star}(R)$ over the stellar half-light radius, $R_{50}$, but not
beyond.  This limit is imposed by the sparsity of reliable $\sigma_{\star}$
measurements for $R\ga R_{50}$ (Martinsson et al.\ 2013; Falc\'{o}n-Barroso
et al.\ 2017; Mogotsi \& Romeo 2019).  Using radial profiles of
$\sigma_{\star}$ derived from CALIFA observations, Romeo \& Mogotsi (2018)
showed that $\sigma_{\mathrm{gg}}(\star)\approx\mbox{0.2 dex}$.  This result
is confirmed by an independent analysis that makes use of scaling relations
(see again Romeo \& Mogotsi 2018), and strengthens the conclusion that the
galaxy-to-galaxy scatter of $\mathcal{A}_{i}$ is smaller for stars (0.1--0.2
dex) than for molecular and atomic gas (0.3 dex).

\subsection{Generating a new set of scaling relations}

The small galaxy-to-galaxy scatter of $\mathcal{A}_{i}$ means that the median
of $\mathcal{A}_{i}$ over the galaxy sample provides a reliable estimate of
the value of $\mathcal{A}_{i}$ in each galaxy:
$\mathcal{A}_{i}\approx\mathcal{A}_{\mathrm{med}\,i}$.  This relation is more
far-reaching than it looks.  In fact, replacing $\mathcal{A}_{i}$ with the
right-hand side of Eq.\ (2) and dividing by $\mathcal{A}_{\mathrm{med}\,i}$,
we get:
\begin{equation}
\frac{j_{i}\hat{\sigma}_{i}}{GM_{i}}\approx1\,,
\end{equation}
where
$\hat{\sigma}_{i}\equiv\overline{\sigma}_{i}/\mathcal{A}_{\mathrm{med}\,i}$.
This is the \emph{key} equation of our paper.  It is not a marginal stability
condition.  It is instead a simple formula that one can use for generating
new scaling relations, all driven by disc gravitational instability.  For
instance, multiply the left- and right-hand sides of Eq.\ (5) by
$M_{i}/M_{\mathrm{ref}}$, and you will get a set of scaling relations for the
relative mass content of each component, where $M_{\mathrm{ref}}$ is any
well-defined reference mass.  Since each scaling relation is derived from
Eq.\ (5) by multiplication, its (logarithmic) scatter will be identical to
the scatter of Eq.\ (5), which itself is identical to the galaxy-to-galaxy
scatter of the parent quantity $\mathcal{A}_{i}$.  We will present the new
set of scaling relations in Sect.\ 4.  Here instead we go on explaining how
to make use of Eq.\ (5).

%%%%%%%%%%%%%%%%%%%%%%%%%%%%%%%%%%%%%%%%%%%%%%%%%%%%%%%%%%%%%%%%%%%%%%%%%%%%%
%%%%%%%%%%%%%%%%%%%%%%%%%%%%%%%%%%%%%%%%%%%%%%%%%%%%%%%%%%%%%%%%%%%%%%%%%%%%%
\begin{table}
\caption{The atomic gas (H\,\textsc{i}), molecular gas ($\mbox{H}_{2}$) and
  stellar ($\star$) $C$-factors appearing in Eq.\ (6) for various choices of
  the averaging radius ($R_{\mathrm{av}}$), measured in units of either the
  stellar half-mass/light radius ($R_{50}$) or the B-band isophotal radius at
  25 mag arcsec$^{-2}$ ($R_{25}$).  In the case of molecular gas, the radial
  range is limited by the sparsity of sensitive CO measurements beyond half
  the optical radius.}
\begin{center}
\begin{tabular}{cccc}
\hline
$R_{\mathrm{av}}$  &  $C_{\mathrm{HI}}$  &  $C_{\mathrm{H2}}$  &  $C_{\star}$  \\
\hline
$1.0\,\,R_{50}$  &  0.4  &  0.4  &  1.2  \\
$1.5\,\,R_{50}$  &  0.4  &  0.4  &  1.4  \\
$2.0\,\,R_{50}$  &  0.5  &  ---  &  1.7  \\
$2.5\,\,R_{50}$  &  0.5  &  ---  &  1.9  \\
$3.0\,\,R_{50}$  &  0.6  &  ---  &  2.2  \\
$0.5\,\,R_{25}$  &  0.4  &  0.4  &  1.4  \\
$1.0\,\,R_{25}$  &  0.5  &  ---  &  2.0  \\
\hline
\end{tabular}
\end{center}
\end{table}
%%%%%%%%%%%%%%%%%%%%%%%%%%%%%%%%%%%%%%%%%%%%%%%%%%%%%%%%%%%%%%%%%%%%%%%%%%%%%
%%%%%%%%%%%%%%%%%%%%%%%%%%%%%%%%%%%%%%%%%%%%%%%%%%%%%%%%%%%%%%%%%%%%%%%%%%%%%

\begin{itemize}
\item If reliable $\sigma_{i}$ measurements are available, then one needs to
  know $C_{i}=1/\mathcal{A}_{\mathrm{med}\,i}$.  In our case
  $C_{\mathrm{HI}}=0.5$, $C_{\mathrm{H2}}=0.4$ and $C_{\star}=2.0$ (see
  Fig.\ 2).  But suppose that one chooses other values of the averaging
  radius $R_{\mathrm{av}}$ than those specified in Sect.\ 2.2, for instance
  because $\sigma_{i}$ is measured only within the inner optical disc.  In
  this case look at Table 1, where $C_{i}$ is calibrated for various choices
  of $R_{\mathrm{av}}$ using L08's sample of spirals, and compute
  $\hat{\sigma}_{i}$ as
  \begin{equation}
  \hat{\sigma}_{i}=
  C_{i}\overline{\sigma}_{i}\,,
  \end{equation}
  where $\overline{\sigma}_{i}$ is given by Eq.\ (4).
\item If reliable $\sigma_{i}$ measurements are not available, then Eq.\ (5)
  is still valid provided that $\hat{\sigma}_{i}$ is redefined as
  $\hat{\sigma}_{i}\equiv(\overline{\sigma}/\mathcal{A})_{\mathrm{med}\,i}$,
  the median of $\overline{\sigma}_{i}/\mathcal{A}_{i}$ over the galaxy
  sample.  Once again, we calibrate this quantity using L08's sample of
  spirals:
  \begin{equation}
  \hat{\sigma}_{i}=
  \left\{
  \begin{array}{rll}
   11\;\mbox{km\,s}^{-1}
  &
  & \mbox{if\ }i=\mbox{H\,\textsc{i}}\,, \\
    8\;\mbox{km\,s}^{-1}
  &
  & \mbox{if\ }i=\mbox{H}_{2}\,, \\
  130\;\mbox{km\,s}^{-1}
  & \!\!\!\!\times\,\,(M_{\star}/10^{10.6}\,\mbox{M}_{\odot})^{0.5}
  & \mbox{if\ }i=\star\,.
  \end{array}
  \right.
  \end{equation}
\end{itemize}

Note that \emph{neither} in Eq.\ (6) \emph{nor} in Eq.\ (7) we have assumed
that $\sigma_{i}=constant$.  In fact, as pointed out in Sects 1 and 2.1, we
have used radial profiles of $\sigma_{i}$, and the resulting
$\overline{\sigma}_{i}$ varies from galaxy to galaxy: for L08's sample of
spirals
$\overline{\sigma}_{\mathrm{HI}}\approx9\mbox{--}31\;\mbox{km\,s}^{-1}$,
$\overline{\sigma}_{\mathrm{H2}}\approx7\mbox{--}33\;\mbox{km\,s}^{-1}$ and
$\overline{\sigma}_{\star}\approx40\mbox{--}108\;\mbox{km\,s}^{-1}$.  It is
remarkable that the $\hat{\sigma}_{\mathrm{HI}}$ and
$\hat{\sigma}_{\mathrm{H2}}$ specified in Eq.\ (7), in spite of being
velocity dispersion scales based on mass and specific angular momentum
[$\hat{\sigma}_{i}=(GM/j)_{\mathrm{med}\,i}$], look like observationally
motivated values of $\sigma_{\mathrm{HI}}$ and $\sigma_{\mathrm{H2}}$.  Note
also that the $\hat{\sigma}_{\star}$ specified in Eq.\ (7) incorporates the
approximate scaling $\overline{\sigma}_{\star}\propto M_{\star}^{0.5}$
(Gilhuly et al.\ 2019; Mogotsi \& Romeo 2019), which is measured across
spiral galaxies of type Sa--Sd and stellar mass
$M_{\star}=10^{9.5\mbox{--}11.5}\,\mbox{M}_{\odot}$.  Finally, it may seem
risky to calibrate $C_{i}$ or $\hat{\sigma}_{i}$ using a single
representative galaxy sample.  However, to the best of our knowledge, L08's
sample is the only one that has reliable published measurements for all the
quantities used in this analysis.  We will test the robustness of our
approach in Sect.\ 4.

\section{GALAXY SAMPLES, DATA AND STATISTICS}

To test the robustness of our approach and present the new set of scaling
relations, we analyse 101 galaxies, from massive spirals to dwarf irregulars,
spanning five orders of magnitude in $M_{\star}$, three and a half orders of
magnitude in $M_{\mathrm{HI}}$, and three orders of magnitude in
$M_{\mathrm{H2}}$.  Such galaxies belong to five distinct samples, which we
name and describe below together with the available data.
\begin{enumerate}
\item `Sp\,(L08+)' is the sample analysed in Sect.\ 2.  It contains 12
  spirals of type Sab--Sc from the THINGS, HERACLES and SINGS surveys.  For
  these galaxies there are published measurements of $M_{\mathrm{HI}}$,
  $M_{\mathrm{H2}}$ and $M_{\star}$ (L08), $j_{\mathrm{HI}}$,
  $j_{\mathrm{H2}}$ and $j_{\star}$ (Obreschkow \& Glazebrook 2014),
  $\sigma_{\mathrm{HI}}$ and $\sigma_{\mathrm{H2}}$ (Romeo \& Mogotsi 2017),
  and $\sigma_{\star}$ (L08).
\item `Sp\,(RM18+)' contains 34 spirals of type Sa--Sd from the EDGE-CALIFA
  survey.  For these galaxies there are published measurements of $\langle
  Q_{\star}\rangle_{50}$, the mass-weighted average of $Q_{\star}(R)$ over
  the stellar half-light radius (Romeo \& Mogotsi 2018), $\sigma_{\star}$
  (Mogotsi \& Romeo 2019), $M_{\star}$ and $M_{\mathrm{H2}}$ (Bolatto et
  al.\ 2017).  There is also a compilation of H\,\textsc{i} masses kindly
  provided by Alberto Bolatto and Tony Wong in advance of publication (the
  sources for the spectra are: van Driel et al.\ 2001; Springob et al.\ 2005;
  Courtois et al.\ 2009; Haynes et al.\ 2011; Masters et al.\ 2014; Wong et
  al., in preparation).  For consistency with the analysis carried out in
  Sect.\ 2, we convert stellar masses from the Salpeter initial mass function
  (IMF) assumed by the CALIFA team (Cid Fernandes et al.\ 2013; S\'{a}nchez
  et al.\ 2016) to the Kroupa IMF assumed by L08, i.e.\ we divide $M_{\star}$
  and multiply $\langle Q_{\star}\rangle_{50}$ by 1.6.  We then divide
  $\langle Q_{\star}\rangle_{50}$ by 3.6 to get $\mathcal{A}_{\star}$.
\item `sp\,(L08+)' contains 4 small spirals of type Sc--Scd from the THINGS,
  HERACLES and SINGS surveys.  For these galaxies there are published
  measurements of $M_{\mathrm{HI}}$, $M_{\mathrm{H2}}$ and $M_{\star}$ (L08),
  $j_{\mathrm{HI}}$, $j_{\mathrm{H2}}$ and $j_{\star}$ (Obreschkow \&
  Glazebrook 2014), and $\sigma_{\star}$ (L08).
\item `sp--dw\,(E17)' contains 9 small late-type spirals and 28 dwarf
  irregulars from the WHISP survey.  For these galaxies there are published
  measurements of $M_{\mathrm{HI}}$, $M_{\star}$, $j_{\mathrm{HI}}$ and
  $j_{\star}$ (Elson 2017).
\item `dw\,(B17)' contains 14 dwarf irregulars from the LITTLE THINGS survey.
  For these galaxies there are published measurements of $M_{\mathrm{HI}}$,
  $M_{\star}$, $j_{\mathrm{HI}}$ and $j_{\star}$ (Butler et al.\ 2017).
\end{enumerate}
Note that for most of the galaxy samples there are no published measurements
of $\sigma_{i}$, hence for those samples $\hat{\sigma}_{i}$ is computed from
Eq.\ (7), as explained in Sect.\ 2.3.  So, while we have shown that velocity
dispersion plays an important role in our approach (see Sect.\ 2), we will
not be able to quantify how significantly it affects the new set of scaling
relations.

%%%%%%%%%%%%%%%%%%%%%%%%%%%%%%%%%%%%%%%%%%%%%%%%%%%%%%%%%%%%%%%%%%%%%%%%%%%%%
\begin{figure}
\includegraphics[scale=.98]{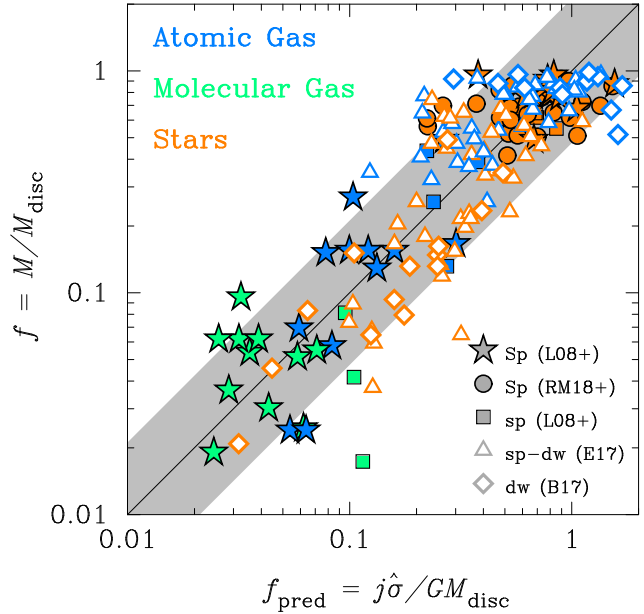}
\caption{Measured ($f$) versus predicted ($f_{\mathrm{pred}}$) mass fractions
  of atomic gas, molecular gas and stars for L08's sample of spirals (star
  symbols), and for the other galaxy samples described in Sect.\ 3.  Also
  shown are the predicted set of scaling relations,
  $f=j\hat{\sigma}/GM_{\mathrm{disc}}$ (black line), and the predicted
  $1\sigma$ scatter, 0.23 dex (grey area).  Note that 82\% of the data points
  fall within the grey area, thus highlighting the accuracy of our
  prediction.}
\end{figure}
%%%%%%%%%%%%%%%%%%%%%%%%%%%%%%%%%%%%%%%%%%%%%%%%%%%%%%%%%%%%%%%%%%%%%%%%%%%%%

%%%%%%%%%%%%%%%%%%%%%%%%%%%%%%%%%%%%%%%%%%%%%%%%%%%%%%%%%%%%%%%%%%%%%%%%%%%%%
\begin{figure*}
\includegraphics[scale=.95]{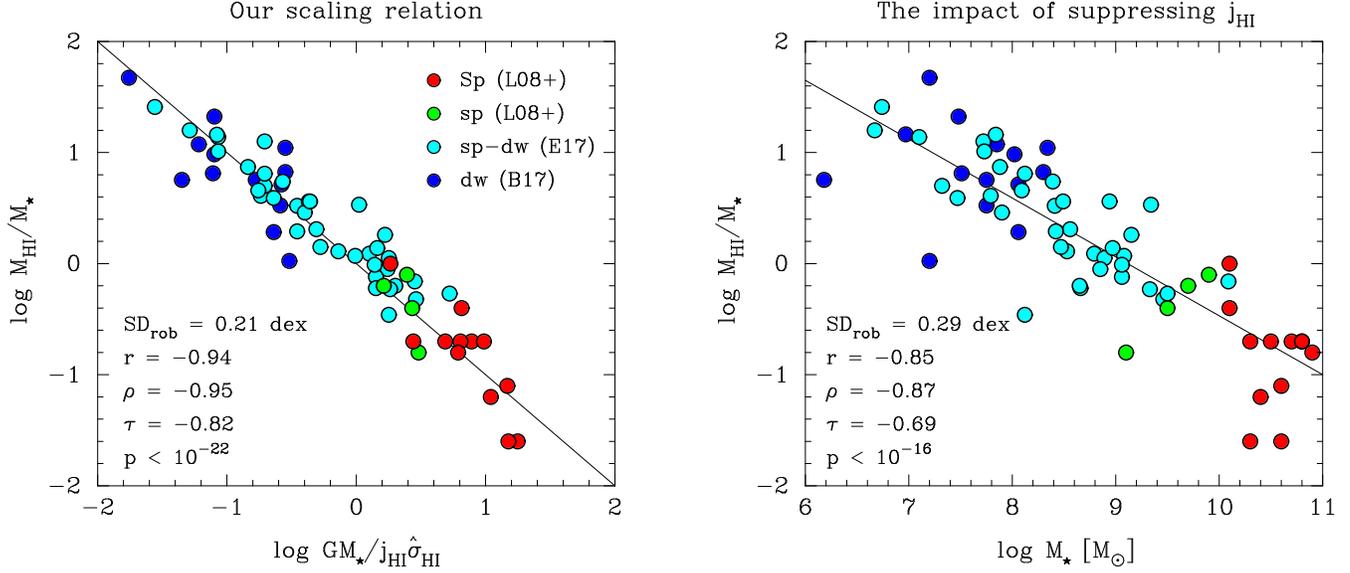}
\caption{Our scaling relation for the relative mass content of atomic gas,
  and the impact of suppressing $j_{\mathrm{HI}}$ (this is the popular
  $M_{\mathrm{HI}}/M_{\star}$ versus $M_{\star}$ scaling relation).  Galaxy
  samples and statistics are denoted as in Sect.\ 3.  The diagonal line in
  the left panel is the prediction based on disc gravitational instability,
  while the line in the right panel,
  $\log(M_{\mathrm{HI}}/M_{\star})=-0.53\log M_{\star}+4.83$, is a robust
  median-based fit to the data points.}
\end{figure*}
%%%%%%%%%%%%%%%%%%%%%%%%%%%%%%%%%%%%%%%%%%%%%%%%%%%%%%%%%%%%%%%%%%%%%%%%%%%%%

To quantify the tightness, strength and significance of the correlations
between the quantities defining the new set of scaling relations, we present
the results of several statistical measures and associated tests.  First of
all, we measure the dispersion of the data points around the predicted
scaling relations using robust statistics:
$\mbox{SD}_{\mathrm{rob}}=1/0.6745\,\mbox{MAD}$, where
$\mbox{SD}_{\mathrm{rob}}$ is the robust counterpart of the standard
deviation and MAD is the median absolute deviation (see, e.g., M\"{u}ller
2000).  Small values of $\mbox{SD}_{\mathrm{rob}}$, for instance 0.2 dex,
mean a tight correlation.  Secondly, we measure Pearson's $r$, Spearman's
$\rho$ and Kendall's $\tau$ correlation coefficients, together with their
significance levels $p_{r}$, $p_{\rho}$ and $p_{\tau}$ (see, e.g., Press et
al.\ 1992).  Values of $r,\rho,\tau\approx(-)1$ and
$p_{r},p_{\rho},p_{\tau}\approx0$ mean a strong and significant
(anti)correlation.

\section{THE NEW SET OF SCALING RELATIONS}

\subsection{The set as a whole}

To give a first view of our scaling relations, we multiply Eq.\ (5) by
$f_{i}=M_{i}/(M_{\mathrm{HI}}+M_{\mathrm{H2}}+M_{\star})$, the mass fraction
of component $i$.  Although $M_{\mathrm{HI}}+M_{\mathrm{H2}}+M_{\star}$ is
the total baryonic mass contained in the disc and in the (pseudo)bulge, we
will simply denote it with $M_{\mathrm{disc}}$ since the sample galaxies are
disc-dominated.  The predicted scaling relations are then
$f_{i}=j_{i}\hat{\sigma}_{i}/GM_{\mathrm{disc}}$.  Parametrizing in terms of
$f_{i}$ is not only natural and commonly used (e.g., Obreschkow et al.\ 2016;
Lutz et al.\ 2017, 2018; D\v{z}ud\v{z}ar et al.\ 2019; Murugeshan et
al.\ 2019), but also useful for visualizing the new set of scaling relations
as a whole, in a single plot.

Fig.\ 3 illustrates that this set stretches across two orders of magnitude in
$f_{i}$, and applies not only to spirals of type Sa--Sd but also to dwarf
irregulars.  This is surprising, considering that such scaling relations have
no free parameters and have been predicted analysing a single representative
sample of spirals [Sp\,(L08+)].  Note also that the $1\sigma$ scatter
measured for the whole galaxy sample is the same as the $1\sigma$ scatter
measured for L08's sample of spirals: $\mbox{SD}_{\mathrm{rob}}=\mbox{0.23
dex}$.

Despite being natural and useful, the $f_{i}$ parametrization is not optimal.
All massive spirals have $f_{\star}\sim1$, and all dwarf irregulars have
$f_{\mathrm{HI}}\sim1$.  This saturation does not influence the measurements
of dispersion, which are entirely determined by the dispersion properties of
the parent quantity $\mathcal{A}_{i}$ (see Sect.\ 2.3), but may influence the
measurements of correlation strength and significance.  In view of that, it
is important to test our scaling relations using other parametrizations.  We
do this in Sect.\ 4.2 for the cold gas scaling relation.  A detailed analysis
of the stellar scaling relation is left for future work.

\subsection{Scaling relation for the relative mass content of cold gas}

To test our atomic gas scaling relation, we use another common
parametrization: $M_{\mathrm{HI}}/M_{\star}$, the atomic gas--to--stellar
mass ratio.  Multiplying the $i=\mbox{H\,\textsc{i}}$ component of Eq.\ (5)
by this quantity, we get the predicted scaling relation,
$M_{\mathrm{HI}}/M_{\star}=
j_{\mathrm{HI}}\,\hat{\sigma}_{\mathrm{HI}}/GM_{\star}$, which can be
rewritten as $\log(M_{\mathrm{HI}}/M_{\star})=
-\log(GM_{\star}/j_{\mathrm{HI}}\,\hat{\sigma}_{\mathrm{HI}})$.  Expressed in
this form, our scaling relation can be directly compared with the popular
$\log(M_{\mathrm{HI}}/M_{\star})$ versus $\log M_{\star}$ scaling relation,
systematically studied by the teams of GASS (e.g., Catinella et al.\ 2010,
2012, 2018), ALFALFA (e.g., Huang et al.\ 2012; Papastergis et al.\ 2012),
and other surveys (e.g., Cortese et al.\ 2011; Peeples \& Shankar 2011).
More importantly, such a comparison allows quantifying the impact of specific
angular momentum on the atomic gas scaling relation.  In fact, the dependence
on velocity dispersion is nearly negligible: for all galaxy samples except
one [Sp\,(L08+)], $\hat{\sigma}_{\mathrm{HI}}$ is fixed (see Eq.\ 7) since
there are no published measurements of $\sigma_{\mathrm{HI}}$ (see Sects 2.3
and 3).  We have checked that fixing $\hat{\sigma}_{\mathrm{HI}}$ even for
Sp\,(L08+) leaves the statistical measurements unchanged.

%%%%%%%%%%%%%%%%%%%%%%%%%%%%%%%%%%%%%%%%%%%%%%%%%%%%%%%%%%%%%%%%%%%%%%%%%%%%%
\begin{figure*}
\includegraphics[scale=.95]{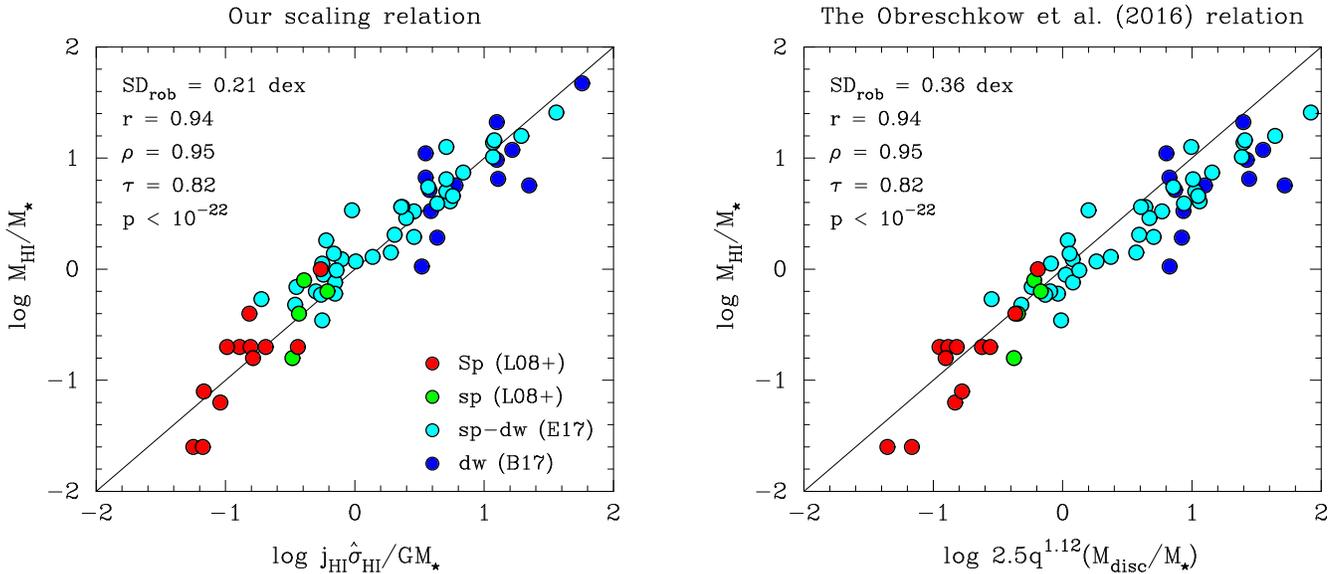}
\caption{Comparison between our scaling relation for the relative mass
  content of atomic gas and the Obreschkow et al.\ (2016) scaling relation,
  where $q=j_{\mathrm{disc}}\,\sigma_{\mathrm{HI}}/GM_{\mathrm{disc}}$ and
  $\sigma_{\mathrm{HI}}=10\;\mbox{km\,s}^{-1}$.  Galaxy samples and
  statistics are denoted as in Sect.\ 3.  The diagonal lines in the left and
  right panels are the two different predictions.}
\end{figure*}
%%%%%%%%%%%%%%%%%%%%%%%%%%%%%%%%%%%%%%%%%%%%%%%%%%%%%%%%%%%%%%%%%%%%%%%%%%%%%

Fig.\ 4 and the statistical measurements shown in the two panels illustrate
that our scaling relation is more constrained than
$M_{\mathrm{HI}}/M_{\star}$ versus $M_{\star}$.  This may seem obvious
because $M_{\mathrm{HI}}/M_{\star}$ versus $M_{\star}$ is one of the gas
scaling relations that have largest scatter (e.g., Catinella et al.\ 2018),%
\footnote{The $M_{\mathrm{HI}}/M_{\star}$ versus $M_{\star}$ relation
  presented here shows significantly less scatter than that presented by the
  GASS team (e.g., Catinella et al.\ 2018), who used a stellar mass--selected
  galaxy sample that is representative in terms of H\,\textsc{i} content.
  This mismatch means that our galaxy sample, which is limited by the
  availability of accurate measurements of specific angular momentum, is not
  fully representative of the H\,\textsc{i} properties of the galaxy
  population in our stellar mass interval.  Obviously, our work shares this
  limitation among all other works that have not used fully representative
  galaxy samples.}
and because it has been demonstrated that $M_{\mathrm{HI}}/M_{\star}$ versus
$M_{\star}$ is `not physical' but driven by the relative fraction of
star-forming and quiescent galaxies as a function of stellar mass (e.g.,
Brown et al.\ 2015).  Note, however, that it is not at all obvious that a
theoretical approach like ours, based on first principles and on a simple
statistical analysis, succeeds in predicting both the slope and the
zero-point of the atomic gas scaling relation across four orders of magnitude
in $M_{\mathrm{HI}}/M_{\star}$, to within 0.2 dex, and without any free
parameter or fine-tuning.  These facts speak clearly.  In particular, the
small scatter of our scaling relation tells us that the mass-averaged
gravitational instability properties of galaxy discs are remarkably uniform
across the sequence Sa--dIrr, and that specific angular momentum has a
significant impact on the atomic gas scaling relation.

A similar conclusion about specific angular momentum was drawn by Obreschkow
et al.\ (2016), so it is interesting to compare our scaling relation with
theirs: $f_{\mathrm{HI}}=\min\{1,2.5\,q^{1.12}\}$, where
$q=j_{\mathrm{disc}}\,\sigma_{\mathrm{HI}}/GM_{\mathrm{disc}}$ and
$\sigma_{\mathrm{HI}}=10\;\mbox{km\,s}^{-1}$.  To be precise, their stability
model predicts that $f_{\mathrm{HI}}=2.5\,q^{1.12}$.  The upper limit on
$f_{\mathrm{HI}}$ was imposed to avoid mass `fractions' $f_{\mathrm{HI}}>1$
for $q>0.44$.  Since this upper limit is not part of the prediction, we do
not consider it.  In addition, since the $f_{\mathrm{HI}}$ parametrization is
open to the criticism pointed out in Sect.\ 4.1, we rewrite the Obreschkow et
al.\ (2016) relation as
$M_{\mathrm{HI}}/M_{\star}=2.5\,q^{1.12}\,(M_{\mathrm{disc}}/M_{\star})$,
which can now be directly compared with ours: $M_{\mathrm{HI}}/M_{\star}=
j_{\mathrm{HI}}\,\hat{\sigma}_{\mathrm{HI}}/GM_{\star}$.  Fig.\ 5 and the
statistical measurements shown in the two panels illustrate that the
Obreschkow et al.\ (2016) relation has the same correlation strength and
significance as ours, but overpredicts $M_{\mathrm{HI}}/M_{\star}$ in dwarf
irregulars significantly, on average by 0.3 dex (a factor of 2).  Our scaling
relation does not show such a bias, and is thus a more reliable atomic gas
tracer.

By analogy with the atomic gas case, our molecular gas scaling relation can
be written as $\log(M_{\mathrm{H2}}/M_{\star})=
-\log(GM_{\star}/j_{\mathrm{H2}}\,\hat{\sigma}_{\mathrm{H2}})$.  This scaling
relation could, in principle, be directly compared with the popular
$\log(M_{\mathrm{H2}}/M_{\star})$ versus $\log M_{\star}$ scaling relation,
systematically studied by the teams of COLD GASS (e.g., Saintonge et
al.\ 2011, 2017), HRS (e.g., Boselli et al.\ 2014), and other surveys (e.g.,
Accurso et al.\ 2017; Gao et al.\ 2019; Liu et al.\ 2019).  Unfortunately, as
shown by Fig.\ 3 and confirmed by Danail Obreschkow (private communication),
there is still a small number of published $j_{\mathrm{H2}}$ measurements,
too few to make a meaningful comparison.  Note, however, that our molecular
gas relation has the same expected scatter as our atomic gas relation (see
Sect.\ 2), so we expect that it should also test well.

\section{CONCLUSIONS}

\begin{itemize}
\item Current models of star formation and/or galaxy evolution commonly
  assume that Toomre's $Q\approx1$.  This assumption is not realistic enough
  to represent the diverse phenomenology of $Q$ in galaxy discs (see
  Fig.\ 1).  The data are now at the point where it is possible to do
  statistics, and look at how much variance there is in $Q_{\mathrm{HI}}$,
  $Q_{\mathrm{H2}}$ and $Q_{\star}$, both from galaxy to galaxy and within a
  galaxy.  The former type of variance is low for all the components, while
  the latter varies significantly not only from stars to gas but also between
  the molecular and atomic gas phases.  The statistical measurements
  presented in this paper impose tight constraints on how self-regulated
  galaxy discs are, and will thus put new-generation models of star formation
  and/or galaxy evolution to a stringent test.
\item The low galaxy-to-galaxy variance of $Q_{i}$ results in a simple
  formula (Eq.\ 5) that one can use for generating new scaling relations, all
  driven by disc gravitational instability.  Eq.\ (5) is the key equation of
  our paper.  We explain how to make use of it in Sect.\ 2.3.
\item Making use of Eq.\ (5), we have generated a new set of scaling
  relations for the relative mass content of atomic gas, molecular gas and
  stars.  We have analysed the set as a whole and each scaling relation using
  more than 100 galaxies, from massive spirals to dwarf irregulars, thus
  spanning five orders of magnitude in $M_{\star}$ and more than three orders
  of magnitude in $M_{\mathrm{HI}}$.  Such tests demonstrate that our scaling
  relations perform well (see Figs 3--5).  In particular, our atomic
  gas--to--stellar mass relation is more physically motivated and has less
  scatter than the Obreschkow et al.\ (2016) relation (see Fig.\ 5).  The
  performance of our scaling relations does not depend on any free parameter
  or fine-tuning, but results from our unified approach for cold gas and
  stars, which robustly predicts not only the scaling but also the zero-point
  of the relations.
\item The tests carried out in this paper also demonstrate that the
  mass-averaged gravitational instability properties of galaxy discs are
  remarkably uniform across the sequence Sa--dIrr, and that specific angular
  momentum plays an important role in such a scenario (see Fig.\ 4).
  Velocity dispersion is another key quantity in our approach.  However,
  limited by the sparsity of published velocity dispersion measurements, we
  have not been able to quantify how significantly it affects the new set of
  scaling relations.
\end{itemize}

Now that we have pointed out the strength of our approach, let us mention its
weakness.  Like all other approaches based on disc gravitational instability,
our approach is calibrated/tested using `disc-dominated' galaxies.  While
this concept is in common usage, observationally it is not well defined
because it says little about the actual bulge mass fraction, B/T, which is an
important `second parameter'.  In view of that, one should test our scaling
relations further, using a more complete galaxy sample and studying the
outcome as a function of B/T.  That test could show potential biases of our
approach and suggest improvements.  We leave that for future work.

Besides testing our scaling relations further, it would be exciting to
generate new ones using Eq.\ (5).  While in this paper we have chosen
baryonic reference masses ($M_{\mathrm{ref}}=M_{\mathrm{disc}},M_{\star}$),
one could parametrize the stellar scaling relation in terms of
$M_{\star}/M_{\mathrm{halo}}$ and compare it with the popular stellar-to-halo
mass relation (e.g., Vale \& Ostriker 2004; Dutton et al.\ 2010; Leauthaud et
al.\ 2012; Behroozi et al.\ 2013; Moster et al.\ 2013; Rodr\'{i}guez-Puebla
et al.\ 2015; van Uitert et al.\ 2016; Posti et al.\ 2019).  One could also
parametrize in terms of $j_{\star}/j_{\mathrm{halo}}$ and compare with the
stellar-to-halo specific angular momentum relation (e.g., Dutton \& van den
Bosch 2012; Kauffmann et al.\ 2015; Posti et al.\ 2018b, 2019).  We leave
such exciting applications of Eq.\ (5) for future work.

\section*{ACKNOWLEDGEMENTS}

We are very grateful to Alberto Bolatto and Tony Wong for providing us with a
compilation of H\,\textsc{i} masses that they made for galaxies from the
EDGE-CALIFA survey (see the references listed in item ii of Sect.\ 3), and to
Ed Elson for providing us with measurements of mass and specific angular
momentum that he made for galaxies from the WHISP survey (Elson 2017).  In
addition, we are very grateful to Oscar Agertz, Alberto Bolatto,
Fran\c{c}oise Combes, John Forbes, Mark Krumholz, Claudia Lagos, Lorenzo
Posti and Florent Renaud for useful discussions.  We are also very grateful
to an anonymous referee for insightful comments and suggestions, for further
constructive criticism and for encouraging future work on the topic.

\bsp

\label{lastpage}

\end{document}